\documentclass{emulateapj}

\usepackage{apjfonts}

\newcommand{\Mo}{M$_{\odot}$}
\newcommand{\Lo}{L$_{\odot}$}
\newcommand{\co}{C$^{17}$O}

\slugcomment{Based on observations carried out with the IRAM Plateau de Bure Interferometer. IRAM is 
supported by INSU/CNRS (France), MPG (Germany) and IGN (Spain).}

\shorttitle{A rotating disk around AFGL~490} 
\shortauthors{Schreyer et al.}

\begin{document}

\title{A rotating disk around the very young massive star AFGL 490}

\author{K. Schreyer}
\affil{Astrophysikalisches Institut und Universit\"atssternwarte, 
Schillerg\"a{\ss}chen 2--3, D--07745
Jena, Germany} \email{martin@astro.uni-jena.de}

\author{D. Semenov, Th. Henning}
\affil{Max--Planck--Institut f\"ur Astronomie, K\"onigstuhl 17, D--69117 Heidelberg, Germany}
\email{semenov,henning@mpia.de}

\and

\author{J. Forbrich}
\affil{Max--Planck--Institut f\"ur Radioastronomie Bonn, Auf dem H\"ugel 69, 
D-53121 Bonn, Germany} 
\email{forbrich@mpifr-bonn.mpg.de}

\accepted{28.12.05 at ApJL} 

\begin{abstract}
We observed the 
embedded, young  8--10~\Mo\ star AFGL~490  at subarcsecond resolution 
with the Plateau de Bure Interferometer 
in the \co\ (2--1) transition and found convincing  evidence that 
AFGL~490 is surrounded by a
rotating disk. Using two-dimensional 
modeling of the physical and chemical disk    
structure coupled to line radiative transfer, we constrain its 
basic parameters. 
We obtain a relatively high disk mass of 1 \Mo\ 
and a radius of $\sim$ 1\,500 AU.
A plausible explanation for the apparent asymmetry of the 
disk morphology is given.
\end{abstract}


%
\section{Introduction}
%
The study of very young massive stars is of particular importance 
for star formation since it is
not yet firmly established by what process, 
disk accretion or stellar merging, such stars form
\citep*[e.g.,][]{YS02,Dobbsea05}. Previous authors
have demonstrated that massive stars can be
formed via  accretion, leading to the formation of large 
($\la 10\,000$~AU) and massive ($\la 1$~\Mo) circumstellar disks. 
However, most of these stars are located 
more than 2~kpc away, which makes it challenging to  confirm    
{\em unambiguously} 
the presence of accretion disks around single objects with radio
interferometry due to the lack of spatial resolution.

We focus on AFGL 490 -- one of the best-studied young 
stars that are in a transition stage to Herbig~Be stars. 
This very young star
is located nearby
\citep[$1\pm0.3$~kpc,][]{Snellea84}, has a mass of 8--10~\Mo, a 
bolometric luminosity of $\sim 2\times10^3$~\Lo\ 
 \citep[B2--3 star,][]{Panagia73}, 
is still embedded in its
parent molecular cloud \citep[A$_{\rm V} 
\sim 40$~mag,][]{ACK89},
and drives a high-velocity outflow 
\citep{Mitchellea95}.
Interferometric measurements by          
e.g.\ \citet[hereafter Paper~I]{Schreyerea02} 
as well as IR observations of 
\citet*{Campbellea86} and \citet*{Bunnea95}
have 
revealed that the star is located inside an ionized $\sim 100$~AU
cavity cleared up by a fast stellar wind and surrounded by
a $\la 500$~AU disk-like structure, which is enshrouded 
in a 22\,000~AU $\times$ 6\,000~AU
envelope. All these structures are further embedded 
in an even more extended envelope.

In this Letter, we present subarcsecond-resolution  PdBI observations of 
AFGL~490 that prove the existence of a rotating disk around AFGL~490 and 
determine its orientation, size, and mass by using comprehensive physical, 
chemical, and line radiative transfer modeling.

%
\section{Observations with the Plateau de Bure interferometer}
We observed AFGL~490 in \co~(2--1) at 224.714389~GHz using the PdBI in 
its B configuration in December 2003 (baselines 61.5m--330.5m). 
The phase reference center of our measurements 
was $\alpha_{2000}$ = 03$^{\rm h}$ 27$^{\rm m}$ 38$^{\rm s\!\!.}$55, 
$\delta_{2000}$ = $+58\degr$ 46$\arcmin$ 59$\arcsec\!\!.$8, and the 
source velocity was set to $V_{\rm lsr}=-13.4$~km\,s$^{-1}$. For the line
measurements, we used one correlator unit with a total bandwidth of 
20~MHz and 512 channels (velocity resolution of 0.052 km\,s$^{-1}$).
Two other correlator units with a bandwidth of 320~MHz 
each were utilized to measure the continuum.

The bandpass and phase calibration were performed on the objects 3C454.3 
and 2145+067. The additional calibrations of the phase and amplitude 
were obtained by observing the objects 0355+508 and 0224+671
every 20 minutes. For the final phase calibration and data reduction 
the Grenoble Software environment GILDAS was applied. 
The maps of $512 \times 512$ square pixels with $0.1\arcsec$ 
pixel size were produced by the Fourier transform of the calibrated 
visibilities using natural weighting. 
The synthesized beam size is
$0.89\arcsec \times0.77 \arcsec$ ($=890~\mathrm{AU} \times 770$~AU 
at 1~kpc) with a positional angle of 73$\degr$. 
For reduction of the continuum, we excluded 
the line emission from the data. Finally, the continuum
was subtracted from the original data in the $uv$-plane,
as it was detected only at the source position.

%
\section{Results}


\subsection{Continuum measurements }
Within the $22\arcsec$ primary PdBI beam at 1.3mm,
AFGL~490 was detected 
point source with a total flux of 1.41~Jy (peak value is 
0.68~Jy\,beam$^{-1}$). 
The continuum is dominated by the thermal 
dust emission because at 1.3mm
the free-free radiation contributes less than 5\% to the
total flux \citep[$\approx 69\pm3$~mJy, see][]{Campbellea86}.

%
\subsection{\co~(2--1) line measurements}

The integrated \co~(2--1) line intensity map overlaid with the 1.3mm
continuum is shown in Fig.~\ref{co}a. 
The continuum is barely resolved at the $5\sigma$ level within
the $\la 2000$~AU area around the star, 
and has a beam-like circular shape. In contrast,
the line intensity map reveals a $\sim 4\,000$~AU arc of emission
that is centered on the continuum peak and has a PA of $\sim$
$15\degr$. Furthermore, the \co\ emission above the 50\% intensity 
level appears as a $\approx$ 700~AU $\times$ 2\,000~AU bar-like 
structure with the main axis  shifted by 400~AU to the
northeast from the center.

The red- and blue-shifted parts of the integrated \co\ (2--1) line 
intensity overlaid with an $H$-band speckle image  
\citep{Alvarezea04} and the 1.3mm
continuum is presented in Fig.~\ref{co}b.  Both lobes
have their maximum 700~AU away from the 1.3mm continuum
point source, and the line connecting
them is shifted 260 AU to the northeast 
(PA 
of the axis perpendicular to the connecting line 
is $\approx 26\degr$).
Due to the simultaneous measurement of both  line and 
continuum in the same bandpass, the positional offset is certainly 
real.  Remarkably, the position of the red- 
and blue-shifted \co\ (2--1) peaks perfectly coincides 
with the position of the CS (2--1) peaks 
reported in Paper~I. The PA of the \co\ emission is smaller 
than the $\sim 45\degr$ positional angle of the 
$20\,000$~AU bar-like structure 
(see Paper~I), 
but  the emission structure
is orthogonal to the large-scale CO outflow 
\citep{Mitchellea95}
and coincides with 
the PA obtained from the infrared polarized map and the 2cm radio emission 
\citep[][]{Campbellea86,haas}.  Note that the \co\ emission
pattern (Fig.~\ref{co}b) is characteristic of outflows or rotating 
disk-like configurations. To distinguish between these two possibilities, we 
analyze the corresponding position-velocity diagram below.

\subsubsection{Disk mass}
We estimate the mass of the AFGL~490 ``inner disk'' from the 
1.3mm continuum emission, using 
Eq.~2 from \citet{Henningea00}.
We assume the standard gas-to-dust mass ratio of 100, a mean 
disk temperature of $100$~K, and a dust mass absorption 
coefficient $\kappa^d_m$ at 
1.3mm between 
1.99   and 5.86~cm$^2$\,g$^{-1}$ 
\citep[grains without icy mantles, 
gas densities 10$^6$ and 10$^8$ cm$^{-3}$, respectively, see][]{OH94}. 
The resulting H$_2$ gas mass is between 0.8 and 2.3~\Mo\ and, thus,
the source-averaged H$_2$ column density is 
$N({\rm H}_2)_{\rm (1.3mm)}= 1 \div 20\cdot10^{23}$~cm$^{-2}$.

The gas mass can be also derived from the \co\ data. 
Due to missing flux of $\sim$30\,\% in the vicinity of AFGL 490, 
even in the reverse map side of the red- and the blue-shifted
line emission, the total integrated flux value is smaller than the sum
of individual contributions from the  red- and blue-shifted emission. 
The sidelobes of the synthesized beam are not contributing to 
the low integrated fluxes due to their low intensities ($<$20\,\%). 
Thus, we focus on the sum of the  red- and the blue-shifted emission.

Assuming optically thin \co\ line emission, the gas mass can be 
derived by 
\begin{eqnarray}
M({\rm H}_2)[{\rm M}_\odot] & = & 1.58\,10^{-10} \ \ 
\frac{ \tau \ (T_{\rm ex}+0.9)  \  {\rm
exp}(16.18/T_{\rm ex}) } { (1-{\rm exp}(-\tau))}      \\     
\nonumber && \times  \ {X({\rm C^{17}O})}
\   ({\rm D[kpc]})2 \int_{line}\!\!\!\!S_{\nu} \Delta{\rm v \ [Jy \ km\,s^{-1}]}  \ ,
\end{eqnarray}
where $\tau=0.01$ is the optical depth of the \co\ (2--1) line, 
$T_{\rm ex}=100$~K is the excitation temperature, 
$X({\rm C^{17}O})=2.5\cdot10^7$ is the H$_2$ abundance 
relative to the amount of C$^{17}$O molecules 
(H$_2$/CO=$10^4$, $^{16}$O/$^{17}$O$=2\,500$), 
and $D=1$~kpc is the distance to the source. 
The masses of the red- and blue-shifted gas are 1.0  and 0.6~\Mo, 
respectively, giving a total value of $\approx1.5$~\Mo\ 
which matches the estimate obtained above.

\subsubsection{Disk velocity field}
The gas velocity distribution across the red- and blue-shifted 
\co\ emission (solid line in Fig.~\ref{co}b) 
is presented in the position-velocity (PV) diagram in 
Fig.~\ref{vco}. The PV map shows a number of single gas 
clumps, indicating an inhomogeneous disk structure. This distribution is 
not caused by optical depth effects as demonstrated
by  radiative transfer calculations (see Sect.\ 3.3).
To fit the PV diagram, we adopt a simple model of a 
Keplerian disk with a mass that linearly increases with radius 
\citep{Vogelea85} and a stellar mass of $8\pm1$~M$_{\odot}$. We assume 
a disk radius of 1\,600~AU based on the spatial extent of the \co\ emission 
and a disk mass between 0.5 and 2~\Mo. With this model, we fit the inner 
clumps ($R < \pm0\arcsec\!\!.7$) of the the PV map and obtain the best-fit 
inclination angle of $35\degr\pm5\degr$ (Fig.~\ref{vco}, solid curve). The fit to the
most intense clumps in the PV map at $R=\pm0\arcsec\!\!.7$ is not as good.
It gives $i=25\degr\pm5\degr$ and super-Keplerian gas velocities in 
the inner disk.
In contrast to massive disks around active galactic 
nuclei that show a modified Keplerian law with an
exponent of $\sim$ $-0.35$ \citep[e.g.,][]{Kon05}, 
the corresponding curve for  AFGL 490 (Fig.~\ref{vco}, dotted curve) 
would only fit to the outer disk part. However, it does not fit 
the inner clumps.
Furthermore, the four 
most intense clumps in the PV map around $-12.6$~km\,s$^{-1}$ can 
be best fitted by a rotating gas ring with a radius of 700~AU and 
similar $i=35\degr\pm5\degr$ (dashed line). Such a small inclination 
angle is in contrast to the apparent \co\  morphology implying
an inclination angle larger than 60$\degr$ 
    for any circularly-symmetric configuration, e.g.\ a disk
    (see ellipse in Fig.~\ref{co}a). 
    However rotation curves for any disk or ring model with larger 
    inclination angles, $i > 40\degr$, can only be well fitted assuming a 
    low-mass central star, $M_\star \le$ 2 \Mo\, which is definitely not 
    the case for AFGL~490. 
    Thus, we propose that there is a massive, clumpy disk around 
    AFGL~490 with nearly face-on orientation ($i\sim30\degr$) 
    and Keplerian rotation.

\subsection{Disk modeling}
The PdBI \co\ (2--1) spectral map of the AFGL~490 disk is presented in 
Fig.~\ref{c17o_fit}.
We fit these spectra using the ``step-by-step'' modeling approach
\citep[][for details see Semenov et al.\ 2006, in preparation]{Semenovea05}. 
Briefly, we apply  the  2D flared-disk
model of \citet{DD04} to simulate the disk physical structure, a gas-grain 
chemical network to calculate time-dependent abundances, and a 2D line
radiative transfer code of \citet{Urania} to synthesize the \co\ spectra. 
We assume that the disk has the power-law surface density 
$\Sigma(r)=\Sigma_0(r_0/r)^p$ and the velocity profile 
$V(r)=V_0(r_0/r)^s$. The 
dust grains have MRN-like size distribution and the gas-to-dust mass 
ratio is 100. 
Moreover, we assume an age of $\sim$0.1~Myr 
due to the lack of  reliable estimates.
\citet{Church} reported the dynamical 
outflow time of 1.8 10$^4$ yr which is an lower limit 
\citep[see][]{Henningea00}.
The best fit to the observed \co\ (2--1) lines 
is obtained with the following disk parameters: (1)  inclination 
and positional angles of $\approx 30\degr$, (2)  inner and outer radii of 
$\la 400$~AU and $1\,400$~AU, respectively, (3) a surface density 
gradient $p\sim -1.5$, and (4) a Keplerian-like velocity  
profile, $s\approx -0.5$.
The best-fit disk mass ranges between $\sim 0.2$ and
1~\Mo. Note that the modeled line profiles
are indeed optically thin, $\tau\la 0.01$. All these values are 
in a good agreement with the observational 
results, and the values derived from the PV diagram.

\section{Discussion and Conclusions}
\begin{figure}
\includegraphics[angle=-90,width=1.0\textwidth,clip=]{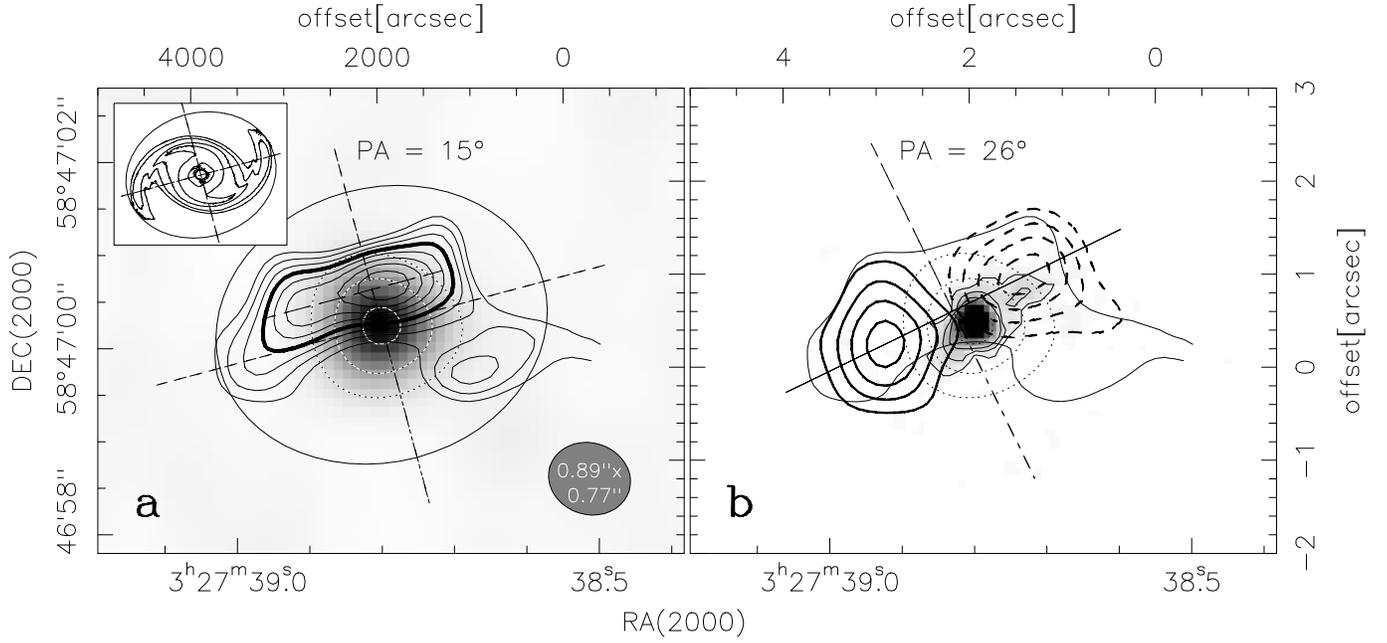}
\figcaption{ 
{\bf (a)} 
Map of the total integrated \co\ (2--1) intensity (thin lines) 
overlaid with the 1.3mm continuum (gray image with dotted lines). 
The contours of the integrated \co\ emission cover intensity levels 
from 20\% ($1.5\sigma$ rms) to 90\% of the peak value 
(0.25 Jy\,km\,s$^{-1}$~beam$^{-1}$) in 10\% 
steps;  the 50\% contour is shown by a thick line. The contour lines
of the continuum emission correspond to 
20\% ($3\sigma$ rms), 50\%, and 90\% of the peak value 
(0.6~Jy\,beam$^{-1}$). 
{\bf (Top left)} 
This chart shows the midplane density distribution of a
self-gravitating massive disk  \citep[Fig.\ 9]{Fromangea04}
inclined to $35\degr$. 
The ellipse indicates the sky area     
covered by such an inclined disk.
{\bf (b)} 
Integrated red- and blue-shifted \co\ line emission overlaid with the 
lowest contour                  
as shown in the left panel. For the blue-shifted lobe, 
solid thick contours correspond to gas with velocities 
between $-15.5$ and $-13.4$~km\,s$^{-1}$, 
while for the red-shifted lobe, dashed-thick contours indicate 
gas with velocities between $-12.5$ and $-9.5$~km\,s$^{-1}$. 
Levels are 30, 50, 70, \& 90\% of the emission peaks  
(blue:  0.09 Jy\,beam$^{-1}$\,km\,s$^{-1}$ ; 
red: 0.12 Jy\,beam$^{-1}$\,km\,s$^{-1}$).
The gray area is the $H$-band speckle 
image from \citet{Alvarezea04}. The straight solid line indicates 
the cut for the position-velocity diagram shown in Fig.~\ref{vco}. 
The dashed lines show the major and the minor axes of the 
emission distributions in the sky plane, whereby the minor 
axis with different position angles (PA) 
is assumed to be in line with the outflow axis. 
\label{co}}
\end{figure}

\begin{figure}
\includegraphics[angle=-90,width=0.8\textwidth,clip=]{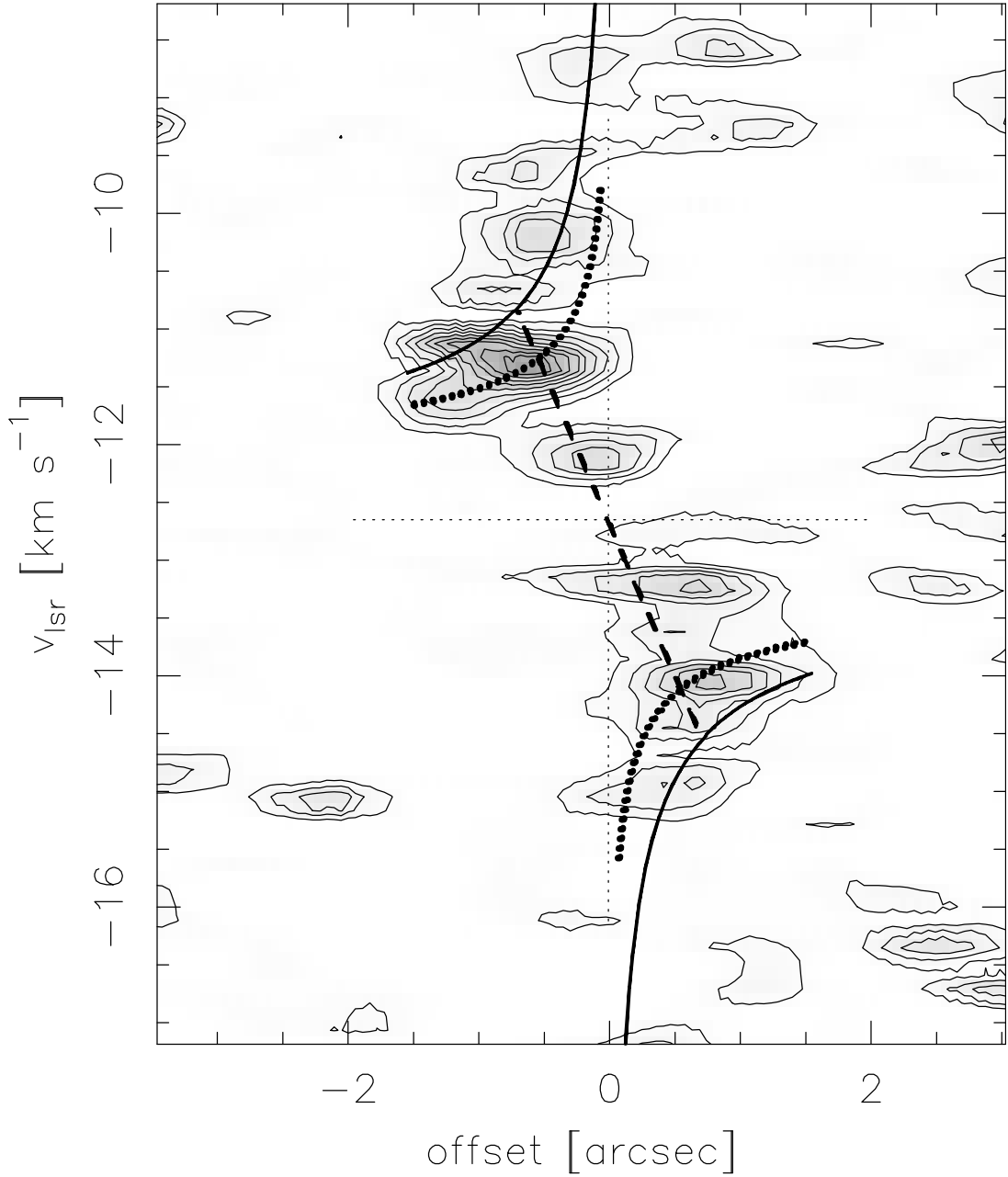}
\caption{The position-velocity diagram of the \co\ (2--1) emission along the 
straight line shown in Fig.~\ref{co}b.    
The thick solid line 
represents the Keplerian disk model with M$_{\rm disk}=1$~\Mo\ 
and $i=35\degr$, while the dashed line corresponds to a model of 
a $700$~AU gas ring with the same orientation. The thick dotted 
curve indicates the disk model with  non-Keplerian rotation, $V(r)
\propto r^{-0.35}$.  
\label{vco}
}
\end{figure}

\begin{figure}
\includegraphics[width=0.65\textwidth,angle=-90,clip=]{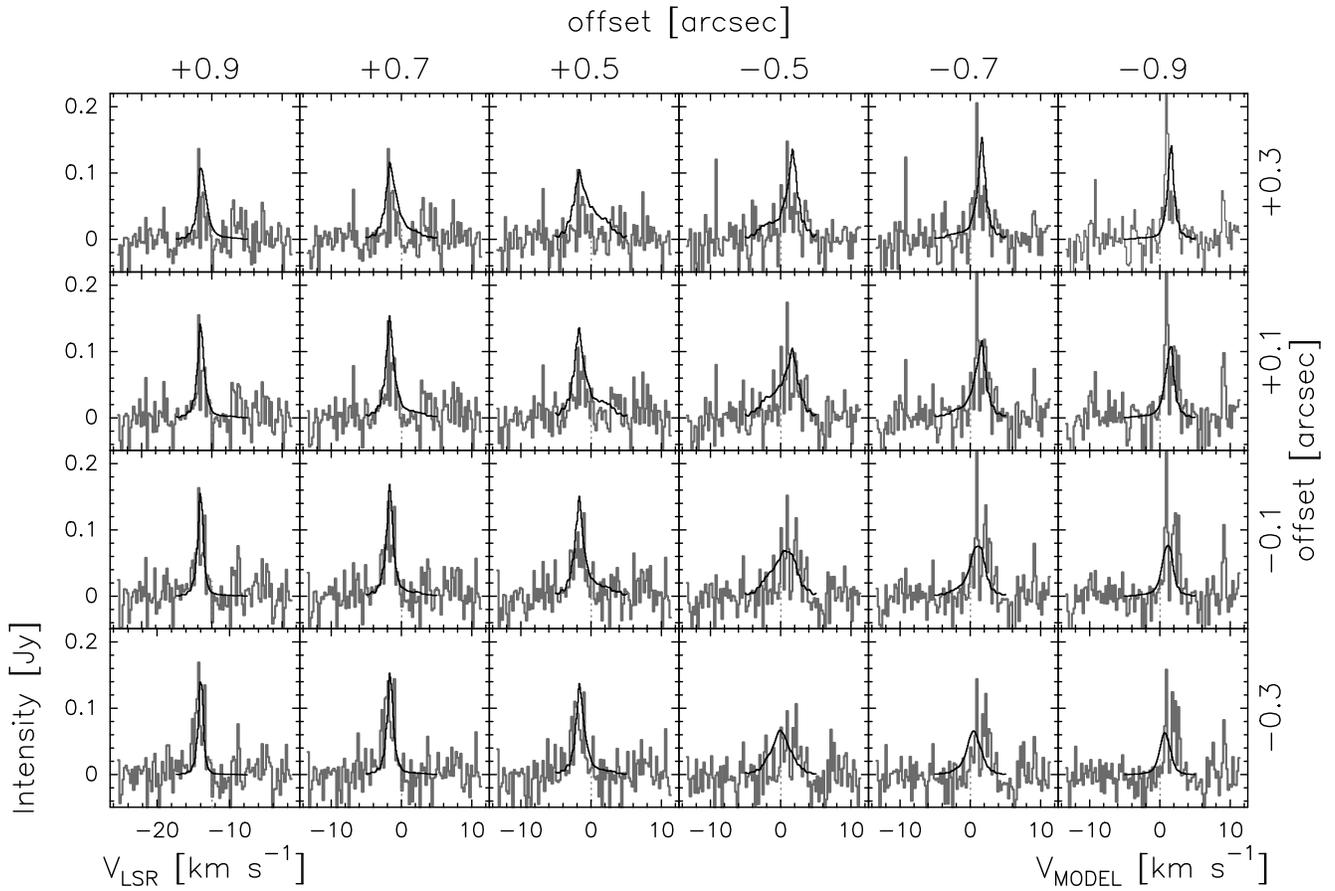}
\caption{The observed PdBI \co\ spectra (thin line) are 
compared to the synthetic ones (thick line) obtained with the best-fit 
model of the AFGL~490 disk. 
\label{c17o_fit}  
}
\end{figure}

Why do we see such a clumpy disk structure? This object 
seems to be relatively young
and could still be in a perturbed state that remains from an earlier, 
non-steady accretion phase. The viscous dissipation timescale 
for the AFGL~490 disk is estimated to be $\sim$ 1~Myr
\citep{Pringle81},  
which far exceeds its age. This hypothesis is further supported by 
the fact that the $\sim 10^4$~years old, large-scale CO outflow
consists of single moving gas clumps, and thus accretion is indeed 
non-steady \citep{Mitchellea95}. Another 
explanation could be that the \co\ (2--1) emission traces the densest 
parts of the spiral arms that are easily excited in massive disks by 
gravitational instabilities. Indeed, the morphology of the \co\ 
emission resembles the density structure of a self-gravitating disk 
with double arms (upper corner in Fig.~\ref{co}a) calculated by 
\citet{Fromangea04}. Moreover, the numerical simulations by 
\citet*{Fromangea04a} imply that such disks should also develop a dual 
structure composed of {\em an inner thin Keplerian disk} fed by a thicker 
self-gravitating torus with nearly uniform rotation. Finally, a 
clumpy disk structure could be a result of a recent encounter with a
nearby star or due to the gravitational interaction with {\rm a 
wide low-mass companion}. The complex multiple outflow systems 
seen in the close vicinity of AFGL~490 strongly support this idea (see
Fig.~4b in Paper~I).

In this Letter, we present clear evidence that $\sim10$~\Mo\ stars 
can be surrounded by Keplerian disks in their earliest 
evolutionally  
phase \citep[see also][]{shepherd}. In contrast to the previously 
reported huge disks around such stars with radii of $\la 10\,000$~AU 
(e.g., Chini et al.\ 2004, but see Sako et al.\ 2005)
the AFGL~490 disk is much smaller, 
$R\sim 1\,500$~AU.
Using advanced theoretical modeling, we constrain basic disk parameters:
(1) the inclination and positional angles are $\sim 30\degr$, 
(2) the surface density profile is $\sim-1.5$, 
(3) the mass is 
$\sim$ 1.5~\Mo\ ($\approx 50\%$ uncertainty), 
and (4)  the disk rotation is close to  Kepler's law.

\begin{acknowledgements}                             
We acknowledge the help of the IRAM staff both of the Plateau de Bure and Grenoble.
\end{acknowledgements} 
%


\clearpage

\clearpage

\clearpage

\end{document}